\documentclass[./sn-mathphys]{sn-jnl}
\usepackage{color}
\usepackage{epsfig}
\usepackage{graphicx}
\usepackage[utf8]{inputenc}

\begin{document}

\def\dert#1#2{ { {\rm d}#1    \over {\rm d}#2 } }
\def\derp#1#2{ { \partial #1  \over\partial #2 } } 
\def\ddt#1{ { {\rm d}\over {\rm d}#1 } }
\def\uvec#1{\hat {\bf e}_{#1}}                       
\def\bfxi{\setbox2=\hbox{$\xi$}
\hbox{{$\xi$}\hskip-.97\wd2
{$\xi$}\hskip-.97\wd2
{$\xi$}\hskip-.97\wd2 {$\xi$}}}
\def\bfalpha{\setbox2=\hbox{$\alpha$}
\hbox{{$\alpha$}\hskip-.97\wd2
{$\alpha$}\hskip-.97\wd2
{$\alpha$}\hskip-.97\wd2 {$\alpha$}}}
\def\bfbeta{\setbox2=\hbox{$\beta$}
\hbox{{$\beta$}\hskip-.97\wd2
{$\beta$}\hskip-.97\wd2
{$\beta$}\hskip-.97\wd2 {$\beta$}}}
\def\bfOmega{\setbox2=\hbox{$\Omega$}
\hbox{{$\Omega$}\hskip-.97\wd2
{$\Omega$}\hskip-.97\wd2
{$\Omega$}\hskip-.97\wd2 {$\Omega$}}}

\def\rd{{\rm d}}                               

\def\spose#1{\hbox to 0pt{#1\hss}}
\def\lta{\mathrel{\spose{\lower 3pt\hbox{$\mathchar"218$}} 
     \raise 2.0pt\hbox{$\mathchar"13C$}}}
\def\gta{\mathrel{\spose{\lower 3pt\hbox{$\mathchar"218$}} 
     \raise 2.0pt\hbox{$\mathchar"13E$}}}

\def\pref#1{\protect{\ref{#1}}}
\def\av#1{\langle #1\rangle}
\def\avU{\langle {\bf U}\rangle}
\def\avB{\langle {\bf B}\rangle}
\def\avA{\langle {\bf A}\rangle}
\def\avJ{\langle {\bf J}\rangle}
\def\turbU{{\bf u}^\prime}
\def\turbB{{\bf b}^\prime}
\def\turbA{{\bf a}^\prime}
\def\turbJ{{\bf j}^\prime}
\def\Ro{{\rm Ro}}
\def\Rm{{\rm Rm}}
\def\urms{u_{\rm rms}}
\def\marginnote#1{\marginpar{\textcolor{blue}{\footnotesize{#1}}}}

\raggedbottom

\title[Evolution of Solar and Stellar Dynamo Theory]{Evolution of Solar and Stellar Dynamo Theory}

\author*[1]{\fnm{Paul} \sur{Charbonneau}}\email{paul.charbonneau@umontreal.ca}

\author[2,3,4]{\fnm{Dmitry} \sur{Sokoloff}}\email{sokoloff.dd@gmail.com}
\equalcont{These authors contributed equally to this work.}

\affil*[1]{\orgdiv{Physics Department}, \orgname{Universit\'e de Montr\'eal}, \orgaddress{\street{C.P.~6128 Centre-Ville}, \city{Montr\'eal}, \postcode{H3C-3J7}, \state{Qc}, \country{Canada}}}

\affil[2]{\orgdiv{Physics Department}, \orgname{Moscow State University}, \orgaddress{\street{Leninskiye Gory}, \city{Moscow}, \postcode{119991}, \country{Russia}}}
\affil[3]{\orgname{IZMIRAN}, \orgaddress{\street{4 Kaluzhskoe Shosse, Troitsk}, \city{Moscow}, \postcode{108840}, \country{Russia}}}
\affil[4]{\orgname{Moscow Center of Fundamental and Applied Mathematics}, \orgaddress{\city{Moscow}, \postcode{119991}, \country{Russia}}}


\abstract{
In this paper, written as a general historical and technical introduction 
to the various review papers collected in the special issue
``Solar and Stellar Dynamo: A New Era'',
we review the evolution and current state of dynamo theory and modelling, with emphasis on the solar dynamo. Starting with a historical survey, we then focus on a set of ``tension points'' that are still left unresolved despite the remarkable progress of the past century. In our discussion of these tension points we touch upon the physical well-posedness of mean-field electrodynamics; constraints imposed by magnetic helicity conservation; the troublesome role of differential rotation; meridional flows and flux transpost dynamos; competing inductive mechanisms and Babcock-Leighton dynamos; the ambiguous precursor properties of the solar dipole; cycle amplitude regulation and fluctuation through nonlinear backreaction and stochastic forcing, including Grand Minima; and the promises and
puzzles offered by global magnethydrodynamical numerical simulations of convection and dynamo action. We close by considering the potential bridges to be constructed between solar dynamo theory and modelling, and observations of magnetic activity in late-type stars. 
}


\maketitle

\section{From first ideas to contemporary state\label{sec:intro}}

The history of solar dynamo studies can be said to begin with the famous talk delivered by J.~Larmor \cite{L19} (the paper is also reprinted in \cite{RSS88}). He attracted the attention of the astronomical community to the fact that the only visible way to obtain solar magnetic fields, as observed a few years before by G.E.~Hale and his colleagues seems to be electromagnetic induction in moving electrically conducting solar media. The car engine was at that time the latest impressive achievement of human civilization, and the idea became known as dynamo theory, after a part of this engine.

Formally speaking, the solar dynamo is an example of a wide variety of various instabilities interesting for astrophysics. Naively speaking, a century may seem as providing sufficient time to investigate an instability in all important details. Dynamos give an interesting and instructive impression here. Initially Larmor's idea proved to be very rich and to contain great potential for development. After various modifications, generalizations and improvements and being  fruitfully combined with other scientific ideas, the dynamo still retains until now quite a lot of problems deserving clarification and development in the context of contemporary science. Our aim here is to describe some of these problems, and present the development of dynamo studies which lead to this new perspective. Of course, we can give here a number of instructive features only, and do not pretend to offer comprehensive review.

A quite obvious point is that Larmor spoke about the Sun rather than say about spiral galaxies just because the concept of galaxies was developed in the next decade only. Application of dynamo theory to the Earth, planets, stars, galaxies, clusters of galaxies, accretion discs, etc, has become an important  part of astrophysics and has offered an attractive and important perspective for exoplanets studies (e.g. \cite{RH04, BrandenburgSubramanian05}). These applications depart from solar dynamo studies in context of specific features of celestial bodies under consideration as well as contemporary observational abilities.

This extensive development of dynamo studies is very important for astronomy, however intensive investigation of the physical basis of solar dynamo looks instructive in a broader scientific context.

First of all, an attempt for straightforward realization of dynamo instability faces the fact that according to Lenz's Law, electromagnetic induction suppresses rather enhances the seed magnetic field. This is why rather simple spherically symmetric or planar 2D flows can not support dynamo action. A number of corresponding antidynamo theorems, from the initial idea of T.~G.~Cowling \cite{Cowling33}  up to sophisticated 2D antidynamo theorem of Ya.~B.~Zeldovich \cite{Z57}  
(see also a further discussion in \cite{ZR80}) were suggested in following decades. Then Yu.~B.~Ponomarenko \cite{P73}  demonstrated that dynamo action is possible in a swirling jet and this idea remains the basis of contemporary experimental dynamo studies (e.g. \cite{LF11, Setal14}). This branch of dynamo studies was able to demonstrate, at the turn of the new millenium, that dynamo is far to be a purely theoretical speculation, but rather a real physical phenomenon which may have in a perspective even practical applications. Laboratory dynamo experiments are quite remote from direct astrophysical applications, however the very laboratory dynamo demonstration is important for solar physics. Contemporary authors do not need to demonstrate how a very weak solar magnetic field can be enhanced up to the kGs values and may admit, if desired, that solar matter was magnetized just in the very early Sun.
The results obtained in these studies also contain an important impulse for theoretical mathematics as a fruitful example of problems for systems of parabolic differential equations with violation of the maximum principle.

Of course, we are interested here in dynamos in astrophysical context. The breakthrough here is associated with the famous migratory dynamo proposed by E.N.~Parker \cite{Parker55}. Parker demonstrated how one can overcome the problem with Lenz law using the idea of two magnetic circuits where the first circuit enhances magnetic field in the second one while the second circuit enhances magnetic field in the first. Conventional dipole magnetic field is associated with the first circuit and is transformed in the magnetic field of the second circuit by solar differential rotation. Magnetic field of the second circuit can be considered as a toroidal magnetic field hidden somewhere in the solar interior. A very important point of the scenario is that the recovery of the poloidal magnetic field from the toroidal one requires mirror asymmetry of the flow.

The point that mirror asymmetry of hydrodynamic flows plays a key role in astrophysical dynamo was suggested at the same time as the idea of effects of P-noninvariance was developed in particle physics (e.g. \cite{LY56}). The interplay between ideas of mirror-asymmetry effects in these two remote domains of physics emerged as a beautiful page of contemporary science.

Parker developed his idea based on his excellent physical intuition. A solid basis for the idea was suggested in a fully  independent research ten year later by physicists from East Germany, namely M.~Steenbeck, F.~Krause and K.-H.~R\"adler \cite{Setal66,Helmbold17} 
(see also \cite{KrauseRaedler80}) in the form of mean-field electrodynamics. The key effect of the scheme responsible for the mirror asymmetry became known as 
the $\alpha$-effect 
(see also Section \ref{sec:mfe} below). 
The idea published initially in an east-German journal and written in German became accessible to the international community practically immediately due to the translation performed by P.~Roberts and M.~Stix \cite{RS71}. Both are well-known experts in the field (e.g. \cite{S04, GR95}) as well as Stix was a native German speaker.

An interplay with ideas and experts from the West and East provides a part of the story very suitable for a novel. In fact Steenbeck, who was a very impressive person, leaves an instructive testimonial in a book about his life \cite{S77}
(see also \cite{Helmbold17}), and some of its elements can be found in novels written later by H.~K\"onigsdorf in years of peaceful revolution in East Germany in the late 1980's. Regarding the Russian side of the story, one of the authors of this chapter keeps in his memory how he passed on to Paul Roberts a proof of the book \cite{Zetal83} on a street just nearby the Kremlin, in what looked like a scene in a spy movie.

Parker \cite{Parker55} as well as Steenbeck et al. \cite{Setal66} considered the initially (almost) unmagnetized medium and associated the mirror asymmetry with Coriolis force action. The magnetic field creates the mirror asymmetry as well, and contemporary understanding of the solar situation is that this contribution is more important. One more point is the importance of meridional circulation as well as other physical effects which understandably were ignored in the early stages of scientific development (more on these in Sections \ref{sec:FTD} and \ref{sec:induction} below). Combined with the ideas of H.~W.~Babcock and R.~B.~Leighton \cite{Babcock61,Leighton64}, this resulted in the contemporary flux-transport model of solar dynamo  (e.g. \cite{C07, DikpatiGilman09}).

The first solar dynamo models dealing with the amplification of a weak seed magnetic field considered the prescribed flow properties (so-called kinematic models). A natural further step was to include a nonlinear dynamo suppression based on some balance arguments and conservation of energy looked as a natural idea for the balance. The situation occurred to be however much more complicated and after the very intensive scientific battle it becomes clear that the magnetic helicity conservation is more important for the problem (viz.~Section \ref{sec:maghelicity} herein). Conservation of magnetic helicity was discovered as early as in the nineteenth century, however nobody considered it as something practically important until K.~Moffatt \cite{Moffatt78} reintroduced the idea in contemporary science. Mathematical aspects of the problem are that magnetic helicities (as well as various other helicities considered in dynamo studies) can be considered as instructive examples of topological invariants and its topological investigation belonged to activities of V.~I.~Arnol'd and his school (e.g. \cite{AK92}).

An important point here is that some crucial dynamo drivers including $\alpha$-effect are associated with topological invariants and being inviscid integral of motions they redistribute in course of dynamo action between various layers in the solar convection shell. Presentation of this redistribution in various solar dynamo models still deserves investigation, however quite a rich bulk of ideas here is accumulated; we mention here as an early achievement the work of Ukrainian astronomer V.~N.~Krividubsky (e.g. \cite{K06}).

Observational identification of dynamo drivers remains a part of astronomy which is still quite remote from its final stage. An important progress here was associated with the idea of N.~Seehafer \cite{S90} who suggested a method to observe magnetic helicity inside sunspots. Due to long-term observations undertaken by Chinese astronomers \cite{Zetal10} time-longitude distribution of magnetic helicity over several solar cycles was observed and the idea propagates on other relevant helicities in further studies by various groups.

Modelling of dynamo action in rotating turbulent spherical shells demonstrated that apart of solar equatorward propagating dynamo waves with polarities which follow the Hale laws, various less convenient magnetic field configurations may be excited (e.g. \cite{JW91}). This may be instructive to explain magnetic activity of some stars and exoplanets.

Helioseismological studies (e.g. \cite{Getal96}) are another (more obvious) way to know more about solar dynamo drivers. Development of helioseismology was associated with one more basic transformation in solar dynamo models (see Section \ref{sec:diffrot} herein).

One more point in dynamo studies to be mentioned here is that the geodynamo models were one of the first cases where direct numerical simulations were able to reproduce very complicated physical processes in various fine details \cite{GR95}. Contemporary solar dynamo models successfully follow this way 
(see Section \ref{sec:MHDsims}). These achievements impressively demonstrated direct numerical simulations and physical explanations of a phenomenon in terms of traditional theoretical physics in two separate problems. The point is that contemporary numerics are so powerful that they can mimic processes which theoreticians still can not explain in traditional terms. It looks as a general challenge for contemporary science to be addressed in its further development.

It is undeniable that in the past century our understanding of astrophysical dynamos has progressed remarkably,
if somewhat non-linearly (in the geometrical sense of the word). Nonetheless, this progress has raised a host of new questions and puzzles, many still outstanding.
The remainder of this review focuses on ``tension points'' left behind by
this meandering path from early ideas to the present state, with emphasis on the solar dynamo.

\section{Tension: Why is mean-field electrodynamics working ?\label{sec:mfe}}

\begin{figure}[t]
\begin{center}
\includegraphics[width=0.7\linewidth]{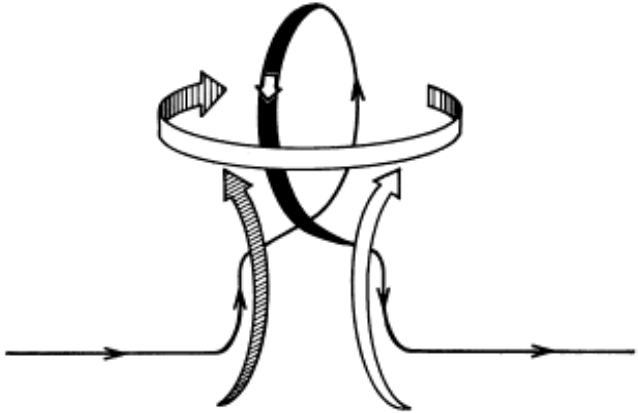}
\end{center}
\caption{
Twisting of an horizontal magnetic fieldline by a cyclonic fluid updraft. In this simple schematic depiction 
the fieldline is twisted outside of the plane of the page, forming a small loop in a plane
perpendicular to the original fieldline. Under the right-hand rule, 
applying Amp\`ere's Law to this small loops yields a current density pointing parallel
to the undeformed magnetic field.
Figure 1 in \cite{Parker70}, reproduced with permission.}
\label{fig:Parker}
\end{figure}

As just discussed,
in the mid-1950's Parker argued 
that the cyclonicity imparted by the Coriolis force
on convective updrafts and downdrafts could effectively break axisymmetry on small spatial scales, and in doing so bypass Cowling's anti-dynamo theorem
\cite{Parker55}. The basic idea is illustrated on Figure \ref{fig:Parker}.
Parker showed that this mechanism could regenerate a poloidal magnetic field from an initially purely toroidal magnetic component, and, operating in conjunction with rotational shearing of
the poloidal component so induced (the so-called $\alpha\Omega$ dynamo scenario), produce a working dynamo loop leading to 
regular polarity reversals.

This groundbreaking idea was soon thereafter formalized through the development of
mean-field electrodynamics \cite
{Steenbecketal66,SteenbeckKrause69} (see also
\cite{Parker70,Moffatt78,KrauseRaedler80,MoffattDormy19}, and references therein).
Separating the flow
and magnetic field into large-scale, slowly varying ``mean'' component 
$\avU,\avB$ and small-scale rapidly varying
``turbulent'' components $\turbU,\turbB$, substitution into the induction equation and averaging
yields an evolution equation for the mean magnetic field:
\begin{equation}
\label{eq:mfmhd}
\derp{\avB}{t}=\nabla\times \left(\avU\times\avB + \bfxi-\eta\nabla\times\avB\right)~,
\end{equation}
where the mean electromotive force $\bfxi$ is given by the average of the small-scale 
flow-field cross-correlation:
\begin{equation}
\label{eq:emf}
\bfxi=\av{\turbU\times\turbB}~.
\end{equation}
Closure is achieved by expanding this turbulent electromotive force (emf) $\bfxi$ in terms of $\avB$ and its derivatives: 
\begin{equation}
\xi_i=a_{ij}\av{B_j}+b_{ijk}\derp{\av{B_j}}{x_k}+...
\end{equation}
This latter expression highlights the fact that mean-field electrodynamics is
fundamentally a linear theory, in the sense that the tensors ${\bf a},{\bf b}$, etc, 
cannot themselves depend on $\avB$, but only on the statistical properties of the turbulent flow.

The symmetric part ($\bfalpha$) of the ${\bf a}$ tensor captures the Parker mechanism of magnetic field deformation by non-mirror-symmetric turbulence, and is now known as the $\alpha$-effect. The three
components of its antisymmetric part can be recast in the form of a pseudo-velocity acting on the mean-magnetic field, called turbulent pumping. 
The antisymmetric part of the rank-3
tensor {\bf b} can be recast as a rank-2 turbulent diffusivity tensor $\bfbeta$ 
\cite{Schrinneretal07}.

The challenge is now to compute these tensorial quantities 
from known statistical
properties of the turbulent flow, which turns out to be a tall order. 
There are three physical regimes under which
this is tractable (see, e.g.,Section 6.3 in
\cite{BrandenburgSubramanian05}; also
Section 3.4.1 in \cite{Ossendrijver03}, 
\cite{Rempel09} and chap.~7 in \cite{MoffattDormy19}).

\begin{enumerate}
\item The energy density of the mean magnetic field is larger than the energy density of the small-scale field;
\item The magnetic Reynolds number is low;
\item The turbulent cyclonic eddies have a lifetime shorter than their characteristic turnover time.
\end{enumerate}

These three physical regimes all amount to the mean magnetic
field suffering little deformation by the small-scale turbulent flow
either because magnetic tension kicks in and prevents large deformation (Regime 1), field/flow slippage occurs and prevents large deformation (Regime 2), or not enough time is available to induce a large deformation (Regime 3). Pictorially,
going back to Figure \ref{fig:Parker}, the magnetic fieldline must be 
twisted out of the plane by an angle $\lta\pi/2$.

Regimes 1 and 2 are most certainly not applicable to solar interior conditions. Regime 3 is harder to assess, as it is notoriously difficult to predict the coherence time of a given turbulent flow, or even to extract it {\it a posteriori}
from a numerical simulation. 
As we shall see presently, some circumstantial evidence exists
suggesting that Regime 3 might be the key.

For turbulence that is 
isotropic and homogeneous, the $\bfalpha$ and $\bfbeta$ tensor reduce to diagonal forms
$\alpha{\bf I}$, $\beta{\bf I}$, with ${\bf I}$ the identity tensor, and
turbulent pumping vanishes.
The second-order correlation approximation then leads to
\begin{equation}
\label{eq:soca}
\alpha=-{1\over 3}\tau_c\av{\turbU\cdot(\nabla\times\turbU)}~,\qquad
\beta={1\over 3}\tau_c\av{(\turbU)^2}~,
\end{equation}
where $\tau_c$ is the coherence time of the small-scale turbulent flow. The $\alpha$-effect
is now simply proportional to the mean kinetic helicity of turbulence, 
and the turbulent diffusivity to its energy density. As shown by
F.~Krause in his Habilitation thesis (as cited in \cite{SteenbeckKrause69}), 
in the
case of a solar/stellar stratified rotating convection zone:
\begin{equation}
\label{eq:Krause}
\alpha=-{16\over 15}\tau_c^2\av{(\turbU)^2}\bfOmega\cdot\nabla (\log{\rho\urms})~,\qquad{\rm [Northern~hemisphere]}
\end{equation}
where $\urms\equiv \sqrt{\av{(\turbU)^2}}$, $\bfOmega$ is the solar
angular velocity vector, and with a sign flip in the Southern solar hemisphere
(see, e.g., Section 6.2 in \cite{BrandenburgSubramanian05}).
Equation (\ref{eq:Krause}) implies that if the properties of turbulence
are independent of latitude, then $\alpha$ is positive in the
Northern solar hemisphere and proportional to $\cos\theta$,
where $\theta$ is the polar angle
(if in doubt, work through footnote 5 in \cite{BrandenburgSubramanian05}).
Except for $\tau_c$, 
the RHS of these expressions are readily extracted from numerical simulations
upon suitable averaging. The
${\bf a}$ and ${\bf b}$ tensors can also be extracted using a variety of techniques 
\cite{BrandenburgSokoloff02,Racineetal11,Schrinneretal07,Augustsonetal15,Simardetal16,Warneckeetal18,Warneckeetal21,Shimadaetal22}.
The turbulent mean-field coefficients so extracted can then be used as input to a classical mean-field solar dynamo
dynamo model, to ascertain whether the resulting large-scale magnetic field evolution resembles ---or not--- that characterizing the parent MHD simulation. Such tests of internal consistency have been carried out succesfully 
\cite{Simardetal13,Simardetal16,Warneckeetal21}, 
with independent numerical simulations and extraction methods. This suggests that mean-field electrodynamics does properly capture the process of turbulent induction and resulting dynamo action, at least in these MHD numerical simulations, and by extension, hopefully, in the sun and stars as well.

Global MHD simulations of large-scale magnetic cycles can also be used to
validate ---or not--- the analytical expressions (\ref{eq:soca}).
This exercise has been carried out by
\cite{Simardetal16} (among others),
estimating $\tau_c$ by the common recipe
consisting in equating $\tau_c$ with the convective turnover time $\ell/u_t$, where $\ell$ and $u_t$ are
local measures of the density scale height and turbulent velocity.
The spatial distributions
they obtain for $\alpha$ and $\beta$ reconstructed from 
Eq.~(\ref{eq:soca})
match tolerably well those directly extracted from their MHD simulation 
(cf.~their Figs.~2 and 6), except for the global amplitude, 
which are larger by a factor of $\approx 5$ in the reconstructions based on
Eq.~(\ref{eq:soca}). 
The amplitudes can be reconciled provided one assumes that
the coherence time $\tau_c$ is one fifth of the convective turnover time, 
which is consistent
with low coherence time turbulence (Regime 3 above). The generality of this
intriguing result remains to be established.

{\bf To sum up:}
Although tractable only in specific parameter regimes of dubious validity 
in the solar/stellar context,
mean-field electrodynamics adequately captures turbulent induction in MHD simulations of solar convection and dynamo action, and leads to internally consistent spatiotemporal evolution of large-scale magnetic fields. Why it actually works so well remains an open question.

\section{Tension: the troublesome magnetic helicity\label{sec:maghelicity}}

Magnetic helicity is a topological invariant measuring the linkage between magnetic flux
systems \cite{Moffatt78,Berger99,MoffattDormy19}.
With the magnetic field expressible as ${\bf B}=\nabla\times {\bf A}$, the magnetic helicity content ${\cal H}_B$ of a volume $V$ of 
magnetized fluid is given by:
\begin{equation}
{\cal H}_B=\int_V {\bf A\cdot B}\,\rd V~.
\end{equation}
In the absence of a flux of helicity at the bounding surface of the volume $V$,
${\cal H}_B$ evolves according to:
\begin{equation}
\label{eq:helB}
\dert{{\cal H}_B}{t}=-2\mu_0\eta\int_V {\bf J\cdot B}\,\rd V~,
\end{equation}
where ${\bf J}$ is the current density and $\mu_0$ the magnetic permeability.
Magnetic helicity is clearly a conserved quantity in the 
ideal (dissipationless) limit $\eta\to 0$, expressing the (topological)
fact that magnetic fieldlines cannot cross one another.

The solar large-scale magnetic field, associated with the magnetic cycle,
is demonstrably helical. In the context of mean-field electrodynamics
(Section \ref{sec:mfe}),
this helicity is imparted on the large-scale
magnetic field by the
$\alpha$-effect, and is of a sign opposite to the kinetic helicity
of the small-scale turbulent flow (viz.~Eq.~\ref{eq:soca}). As the large-scale
magnetic field is amplified, so must ${\cal H}_B$, in apparent violation
of the above conservation argument. Note that polarity reversals
 of the large-scale magnetic field are irrelevant to the problem; reversing the
magnetic polarity flips the sign of both ${\bf A}$ and ${\bf B}$, leaving the sign of ${\cal H}_B$ unchanged. How, then, can the solar large-scale magnetic
field wax and wane in the course of the magnetic cycle ?

Here again mean-field electrodynamic offers some useful insight. 
Applying scale separation
to the vector potential ${\bf A}$ and current density ${\bf J}$, two evolution
equation for the magnetic helicity associated with the large- and small-scale magnetic components can be obtained \cite{BrandenburgSubramanian05,Rempel09}; in the ideal ($\eta\to 0$) limit:
\begin{eqnarray}
\label{eq:helBls}
\ddt{t}\int_V\avA\cdot\avB\,\rd V &=&
+2\int_V\bfxi\cdot\avB\,\rd V~,\\
\ddt{t}\int_V\av{\turbA\cdot\turbB}\,\rd V &=&
-2\int_V\bfxi\cdot\avB\,\rd V
\label{eq:helBss}
\end{eqnarray}
with the mean electromotive force $\bfxi$ given by Eq.~(\ref{eq:emf}). Observe
that the action of the turbulent emf on the large-scale magnetic field, i.e.,
the terms on the RHS of Eqs.~(\ref{eq:helBls})--(\ref{eq:helBss}),
produces magnetic helicity of opposite signs at large and small scales,
so that the \emph{total} magnetic helicity produced by $\bfxi$ acting on
$\avB$ is thus nil.
The large-scale field $\avB$ can now be amplified because magnetic helicity of
opposite sign builds up at small scales. However, this turns out
to oppose the $\alpha$-effect,
as originally demonstrated by \cite{Pouquetetal76}
(see also Section 9 in \cite{BrandenburgSubramanian05}).
The large-scale field is amplified, but at the cost of rapidly quenching the inductive part of the
turbulent emf \cite[see][]{Kleeorinetal95,BlackmanBrandenburg02,Brandenburgetal09}.

A way out of this quandary was identified in \cite{Brandenburg01}. It
consists in invoking a direct turbulent
cascade of helicity towards even smaller scales than that at which $\bfxi$ is operating,
so that Ohmic dissipation sets in, as per Eq.~(\ref{eq:helB}),
and dissipates the helicity produced
at the inductive scale (RHS of Eq.~(\ref{eq:helBss})). 
At this dissipative scale, the magnetic Reynolds number
$\Rm=u_t L/\eta$ is of order unity, but remains much larger
at the inductive scale of $\bfxi$, and even larger yet at the scale of $\avB$,
so that Eqs.~(\ref{eq:helBls})--(\ref{eq:helBss}) effectively hold.
Now the $\alpha$-effect can operate, and a helical large-scale magnetic field
can grow.

Another mechanism allowing to evade the constraint of magnetic helicity dissipation is to evacuate it through the volume boundaries. A star like the sun is not embedded in a true vaccum, and magnetic
helicity can be evacuated through the corona. In particular,
coronal mass ejections have been suggested to contribute significantly to the global magnetic helicity budget
\cite{BieberRust95,Low01,Greenetal03,Lynchetal05}.
Avoiding $\alpha$-quenching via helicity flux across domain boundaries 
and/or cancellation across the equatorial plane
has also found support in MHD numerical simulations
\cite{BrandenburgDobler01,Warneckeetal11}.

{\bf To sum up:}
Conservation of magnetic helicity in the high-$\Rm$ regimes poses a strong constraint on magnetic field amplification by turbulent induction,
and can potentially quench the growth of the solar large-scale magnetic field.
This constraints can be bypassed by a double turbulent cascade or expulsion
of helicity from the region of dynamo action. Which of these mechanisms (if any or either) is regulating the overall solar magnetic helicity budget, 
remains an open question. 

\section{Tension: The troublesome solar differential rotation\label{sec:diffrot}}
Already in the nineteenth century, R.C.~Carrington and G.~Sp\"orer independently
noted that sunspots emerge closer and closer to the solar equator as the sunspot cycle unfolds.
The first convincing explanation for this striking spatiotemporal pattern was proposed almost a century later by Parker, in the form of dynamo waves \cite{Parker55}. In
$\alpha\Omega$ mean-field dynamos, these waves propagate in a direction 
${\bf s}$ given by
\begin{equation}
\label{eq:dynwaves}
{\bf s}=\alpha\nabla\Omega\times \uvec{\phi}~,
\end{equation}
a result now known as the Parker-Stix-Yoshimura sign rule. 
Extending the observed surface latitudinal differential rotation pattern 
inwards along cylindrical isosurfaces yields a positive radial shear component
at low latitudes, which then requires a negative $\alpha$-effect 
in the Northern hemisphere\footnote{Or more precisely: a negative $\alpha_{\phi\phi}$ tensor component, 
the only component typically retained in classical $\alpha\Omega$ mean-field models.} to
produce equatorward propagation \cite{Yoshimura75,Stix76}.

\begin{figure}[h]
\begin{center}
\includegraphics[width=0.8\linewidth]{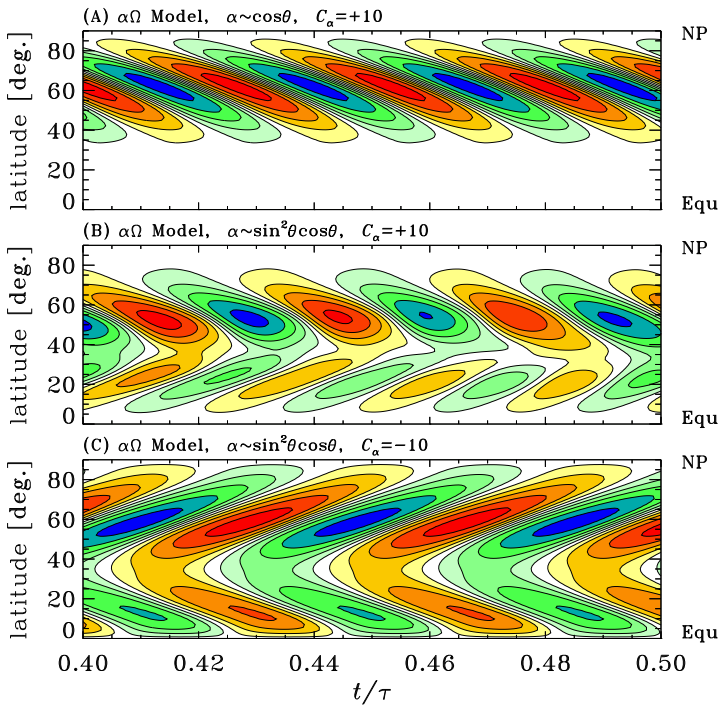}
\end{center}
\caption{
Northern hemisphere time-latitude diagrams of the large-scale toroidal magnetic component for
three mean-field kinematic axisymmetric classical $\alpha\Omega$ dynamo solutions, all using the same solar-like parametrization of the solar internal differential rotation, but different
latitudinal profile and sign for the $\alpha$-effect, in all cases concentrated near the base
of the convective envelope ($r/R=0.7$, where the diagram are constructed). 
The toroidal magnetic field are normalized to their peak amplitude, 
and isocontours are equally spaced, with yellow$\to$red (green$\to$blue) indicating positive (negative) values. 
Figure 7 in \cite{Charbonneau20}, used with permission.}
\label{fig:3bfly}
\end{figure}

This neat picture was thrown into disarray by the first helioseismic inversions of the solar internal differential rotation \cite{Brownetal89,Dziembowskietal89}. 
Rather that cylindrical isocontours
of angular velocity, these inversions revealed that the surface differential rotation remains constant along approximately
radial segments, yielding a shear that is primarily latitudinal within the bulk of the convection zone, transiting beneath it to near-solid body rotation across a thin rotational
shear layer since known as the tachocline \cite{SpiegelZahn92,Tomczyketal95}. 
As shown on Figure \ref{fig:3bfly}, this complex form of the solar internal differential rotation yields very complex patterns of dynamo waves, 
even if the $\alpha$-effect is artificially concentrated at the base of the convection zone, to suppress induction by the purely latitudinal shear above. 
The Figure shows Northern hemisphere
time-latitude (``butterfly'') diagrams for the toroidal magnetic component
at the base of the convection zone, using different latitudinal dependency and sign
for the $\alpha$-effect in a classical $\alpha\Omega$ mean-field model
(for more on these dynamo solutions, see Section 4.2 in \cite{Charbonneau20}).
Here, and even with the $\alpha$-effect concentrated towards low latitude via a 
$\sin^2\theta\cos\theta$ dependency on polar angle (panels B and C), 
the strong positive radial shear in the high latitude regions of the tachocline
dominate induction, leading to multiple branches and activity peaking at much 
higher latitudes than observed.

Within the standard $\alpha\Omega$ dynamo modelling framework, the only way to achieve equatorward dynamo wave propagation is to strongly concentrate a (negative) $\alpha$ effect not only radially at the base of the convection zone, but also latitudinally in its equatorial regions.
Recall from Eq.~(\ref{eq:Krause})
that the minimal latitudinal dependency expected from cyclonic 
convection leads to a positive $\alpha$-effect concentrated at high latitudes,
with $\alpha\propto\cos\theta$; 
this does lead to equatorward propagating dynamo waves (viz.~Fig.~\ref{fig:3bfly}A), but again peaking at far higher
latitudes than observed on the Sun.

Helioseismology 
has also revealed the presence of a thin subsurface radial shear layer
extending from the equator to mid-latitudes;
equatorward propagating dynamo waves concentrated 
at low latitudes can then be produced in conjunction with a positive Northern hemisphere
$\alpha$-effect, but the
small thickness of the layer sets the 
length scale of dynamo eigenmodes, typically leading to 
multiple overlapping magnetic flux systems even for weakly supercritical dynamos.

{\bf To sum up:}
in classical $\alpha\Omega$ mean-field dynamo models
built using 
the solar internal differential rotation profile as inferred from helioseismic inversions, 
it is very hard to produce 
a sunspot butterfly diagram-like dynamo wave propagation pattern without making some very 
{\it ad hoc} assumptions regarding the spatial distribution and/or sign of the turbulent $\alpha$-effect.

\section{Tension: Flux transport by meridional flows\label{sec:FTD}}

Equatorward propagation of activity belts in the course of the cycle can also be achieved through advection by a bulk meridional flow acting 
as a ``conveyor belt'' displacing the internal
toroidal magnetic field equatorward as it is amplified
by rotational shearing \cite{Wangetal91,Choudhurietal95,Kuekeretal01,Rempel06a,PipinKosovichev11}. 
Dynamo models achieving a solar-like butterfly diagram in this manner are known as \emph{flux transport dynamos}
(see \cite{DikpatiGilman09,Karaketal14} for dedicated reviews). 
Adding a meridional flow to a classical $\alpha\Omega$ dynamo model of the type considered in Section \ref{sec:diffrot}, one finds that bulk transport of the magnetic field overwhelms the dynamo wave provided advection by the meridional flow dominates
over diffusive transport. The ratio of these two transport mechanisms is quantified via a magnetic Reynolds number:
\begin{equation}
\label{eq:magRe}
\Rm={u_0 L\over \eta_T}~,
\end{equation}
where $u_0$ is a typical speed for the meridional flow, $\eta_T$ is a turbulent magnetic diffusivity, and $L$ is a characteristic length scale for the meridional flow, 
usually taken as the solar radius. 
For magnetic flux transport to take place in the desired manner, 
this magnetic Reynolds number must be relatively high, i.e., $\sim 10^2$ or more.
See Section 4.4 in \cite{Charbonneau20} 
for some representative dynamo solutions.

Powered by Reynolds stresses and pole-equator temperature differences caused by rotational influence on convective energy transport,
meridional flows are as unavoidable as differential rotation in a rotating, stratified turbulent
convective envelope 
\cite{Kippenhahn63,Ruediger89,Kitchatinovetal94,MieschToomre09,Balbusetal12,FeatherstoneMiesch15}. 
This flow is observed at the solar surface, poleward-directed and with speeds peaking in the range 10--15 m s$^{-1}$ at mid latitudes, with some variations in phase with the solar cycle \cite{Hathaway96,Ulrich10,CameronSchuessler10}. 
Mass conservation evidently requires an equatorward return flow
somewhere within the convection zone. Helioseismic determinations of the internal meridional flow have yielded conflicting results, some inversions suggesting 
a very shallow equatorward return flow \cite{Jackiewiczetal15},
others a complex flow pattern characterized by multiple flow cells stacked in radius and/or latitude
\cite{Schadetal13,Zhaoetal13}, while others yet are consistent with a single cell per meridional quadrant, 
with the equatorward return flow peaking near the base of the convection zone
\cite{RajaguruAntia15,Gizonetal20}.

Global numerical simulations of stratified rotating convection do provide additional insight on the matter. Solar-like differential rotation, in the sense of the rotation rate decreasing monotonically from the equator towards the poles, materializes when the Rossby number
$\Ro=u_t/2\Omega L$ is sufficiently small, $\lta 0.3$, but in this regime the meridional flow is 
markedly multi-celled.
A single meridional flow cell per meridional quadrant is achieved at higher Rossby number, but the differential rotation is no longer solar, with the equatorial regions rotating more
slowly than the mid-latitudes\footnote{Note however that some mean-field turbulence models do produce a single-cell meridional flow in conjunction with a solar-like differential rotation profile; see, e.g., \cite{KitchatinovOlemskoy11}.}
(\cite{Gastineetal14,Brunetal22}.
in particular their Fig.~10).
Interestingly, given its rotation rate and luminosity, the sun appears to be characterized by a Rossby number near the tipping point between these two regimes, and some global MHD simulations indicate
that magnetic stresses may turn an anti-solar differential rotation into a solar-like profile, while generating a solar-like single-cell meridional 
flow profile
(see \cite{Karaketal15,HottaKusano21,Hottaetal22},
and references therein).

The implications of single vs multi-cell meridional flows for flux transport dynamos 
are profound. Multiple meridional flow cells can lead to a variety of time-latitude patterns departing significantly from the observed sunspot butterfly diagram
\cite{JouveBrun07,PipinKosovichev13,Beluczetal15}. 
The dynamo simulations of \cite{Hazraetal14} do suggest
that if the diffusivity is sufficiently high, an equatorward drift of
the deep toroidal field can be achieved even in the presence 
multiple flow cells stacked in radius, provided the deeper cell has an equatorward return flow at the base of the convective envelope. 
The key parameter then becomes the magnetic Reynolds number (\ref{eq:magRe}), 
which is critically dependent on the assumed value for the (turbulent) magnetic diffusivity,
a notoriously difficult quantity i
to compute from first principles\footnote{
Alternate versions
of flux transport dynamos can be constructed by relying on
turbulent pumping 
(see \S 2) 
to achieve, in part or in its entirety,
downward transport of the surface field
\cite{Jiangetal13} and equatorward drift of the deep toroidal field
\cite{GuerreroDalPino08,HazraNandy16}.
Measurements of turbulent pumping 
in some MHD numerical simulations of solar convection do yield
strong subsurface downward pumping as well as equatorward latitudinal pumping
at mid- to low latitudes within the convecting fluid layers, with speeds of a few meters per second
\cite{Ossendrijveretal02,Racineetal11,Simardetal16,Warneckeetal18,Shimadaetal22},
similar to the deep meridional flow speed.}.

{\bf To sum up:} 
flux transport dynamo can in principle produce solar-like ``butterfly diagrams'' 
even in cases where classical dynamo waves would do otherwise; however,
their proper operation depends sensitively
on the spatial form of the internal axisymmetric meridional flow, as well as on the value of turbulent magnetic diffusivity produced by solar convection.

\section{Tension: Competing inductive mechanisms?\label{sec:induction}}

The turbulent $\alpha$-effect is by no means the only way to evade Cowling's theorem. 
Originally proposed by Babcock \cite{Babcock61} and developed quantitatively by
Leighton \cite{Leighton64,Leighton69}, 
but largely eclipsed by the rise of mean-field electrodynamics
until its vigorous revival a quarter of a century later 
\cite{Wangetal91,Choudhurietal95,Durney95,DikpatiCharbonneau99,NandyChoudhuri01},
what is now known as the Babcock-Leighton mechanism is arguably its most
convincing alternative.

A little over a century ago Hale and collaborators established a number
of empirical Laws describing the cycle-to-cycle variations in the hemispheric
pattern of magnetic polarity measured in sunspots \cite{Haleetal19}.
They also established what is
since known as \emph{Joy's Law} namely the systematic inclinations of the line segment
joining the poles of bipolar sunspot groups with respect to the E-W line, this tilt angle ($\gamma$) increasing with heliocentric latitude ($\lambda$).
Leighton \cite{Leighton69} originally
parametrized this variation as
\begin{equation}
\label{eq:Joy}
\sin\gamma=0.5\sin\lambda~,
\end{equation}
but other related forms fit the data equally well,
in view of the large
scatter of observed tilt angles about such mean relationships
(see, e.g., \cite{McClintockNorton13} and references therein). 

Bipolar magnetic regions (BMRs) are believed to originate from magnetic 
flux ropes buoyantly rising through the convection zone and piercing the photosphere as ``$\Omega$-loops'' \cite{Parker55b}. 
Modelling of this process in the thin flux tube approximation has allowed to identify the physical underpinning
of Joy's Law, in the action of the Coriolis force
on flows developing along the axis of
buoyantly rising magnetic flux ropes 
\cite{DSilvaChoudhuri93,Fanetal93,Caligarietal95},
and/or via the asymmetric buffeting imparted by cyclonic convection
\cite{Weberetal11,Weberetal13}
(see \cite{Fan21} for
a comprehensive review).

\begin{figure}[t]
\begin{center}
\includegraphics[width=0.9\linewidth]{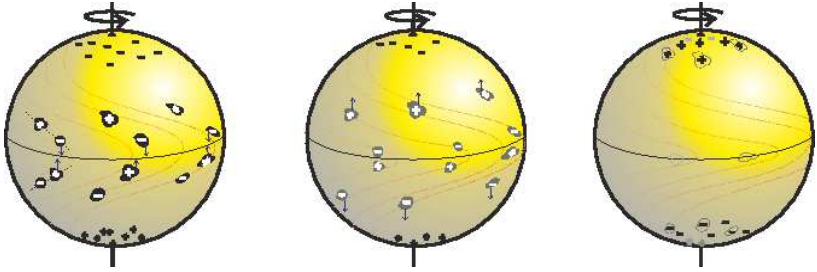}
\end{center}
\caption{
Schematic illustration of the Babcock-Leighton mechanism in operation.
At left, BMRs have just emerged, abiding to Hale's polarity Laws
as well as Joy's Law. After some time (middle), the BMRs have decayed
and spread diffusively, with preferential transequatorial dissipation
of the leading polarities, and transport of the residual 
trailing polarity to high latitudes by surface flows.
This eventually leads to the reversal of the pre-existing 
dipole (here negative), and buildup of a new (positive) dipole (at right).
Diagram produced by D.~Passos, used by permission.}
\label{fig:BLmech}
\end{figure}

The global dipole contribution $\delta D$
associated with a BMR
carrying a magnetic flux $\Phi$ with pole separation $d$ and emerging 
at latitude $\lambda$ 
\cite{Petrovayetal20}
is given by:
\begin{equation}
\label{eq:dipole}
\delta D={3\cos\lambda\over 4\pi R^2}\Phi d \sin\gamma~.
\end{equation}
As BMRs decay and ``release'' this dipole moment, preferential
cross-equatorial dissipation of the leading magnetic 
polarity and transport of the trailing
polarity towards the poles leads to polarity reversal and subsequent
buildup of a new global dipole moment. This surface transport of magnetic flux
is observed at the solar surface, and has been modelled in detail
\cite{Wangetal89,Jiangetal14,UptonHathaway14,Lemerleetal15,Whitbreadetal17}, leaving little doubt to its role in reversing the surface dipole.
Figure \ref{fig:BLmech} illustrates schematically this sequence of events.
Note that the tilt embodied in Joy's Law is crucial here; if BMRs emerge
aligned with the E-W direction ($\gamma=0$), then $\delta D\equiv 0$; 
both poles of the BMR then experience the same cross-equatorial 
diffusive cancellation, leaving behind and no net hemispheric flux.

From end-to-end, the sequence of flux tube formation, destabilization, emergence as BMRs, surface decay and transport, thus converts a positive (negative)
internal toroidal field
into a positive (negative) dipole moment, in a manner
analogous to a positive $\alpha$-effect 
in mean-field electrodynamics\footnote{Note that in both cases, the Coriolis
force ultimately
provides the break of axisymmetry needed to evade Cowling's theorem.}.
Rotational shearing of this large-scale dipole can then regenerate the toroidal component and close the dynamo loop. 

As with mean-field dynamos based on the turbulent $\alpha$-effect, 
Babcock-Leighton dynamos must abide with helicity conservation. 
Magnetographic observations indicate that the magnetic flux ropes emerging as BMRs carry magnetic helicity in the form of internal twist about their axis 
(see, e.g., \cite{Seehafer90,LopezFuentesetal03}), 
as expected if they form from a deep-seated 
dynamo-generated large-scale
magnetic field that is itself helical. Ultimately,
the large-scale twist of the flux rope itself (or writhe) associated with 
Joy's Law, acts as the global source of magnetic helicity 
in this class of dynamos.
For more on these matters, see \cite{Pevtsovetal14} and references therein.

Most contemporary versions of solar cycle models based on the Babcock-Leighton mechanism are formulated as flux transport dynamos, with the meridional flow
carrying the surface dipole to the deep interior\footnote{Turbulent pumping has also been invoked as a flux transport mechanism in this context, in particular
to ensure submergence of the surface magnetic field; 
see, e.g., \cite{Jiangetal13}.},
where
rotational shearing takes place, and driving the equatorward propagation
of emerging BMRs in the course of the cycle (viz.~Section \ref{sec:FTD}). 
It must be emphasized that to operate properly, \emph{all} such solar 
dynamo models must invoke a strongly enhanced magnetic diffusivity, 
presumably of turbulent origin, as provided by mean-field theory.
For more on such models, see Sections 5.4 and 5.5
in \cite{Charbonneau20}.

The polar cap (latitudes $>60$ degrees) magnetic flux amounts to
$\sim 10^{14}\,$Wb at times of peak surface dipole strength. 
The total unsigned magnetic flux emerging in the form of BMRs adds up to $\sim 10^{17}\,$Wb in the course of a typical activity cycle. The toroidal-to-poloidal conversion efficiency of the Babcock-Leighton mechanism
thus needs not be high, of order $\sim 0.1\,$percent only. 
In fact it has been argued
that the poloidal flux generated by the Babcock-Leighton mechanism is indeed sufficient, in conjunction with rotational shearing, 
to account for the emerging magnetic flux 
\cite{CameronSchuessler15}, although turbulent induction in the interior
cannot be ruled out via this argument.
Is the Babcock-Leighton mechanism then essential to the solar magnetic cycle ?
Answering this question on the basis of observations
would require a detailed magnetic flux budget of the solar polar caps, i.e.,
accounting for flux emergence, submergence, transport from lower latitudes,
as well as local generation.

{\bf To sum up:} The Babcock-Leighton mechanism is observed operating
at the solar surface, and in itself can account for the reversal of the
surface dipole. Whether the surface dipole 
so generated feeds back into the dynamo loop, or is a mere side-effect of
a deep-seated turbulent dynamo operating independently in the solar interior,
remains an open question.

\section{Tension: The surface dipole as precursor\label{sec:precursor}}

The surface dipole strength at activity minimum is long known to be a good precursor for the amplitude of the upcoming sunspot cycle
\cite{Schattenetal78,Svalgaardetal05}.
(for reviews of solar cycle prediction schemes, see \cite{Pesnell16,Petrovay20}).
The dipole-as-precursor is also implemented in some dynamo model-based cycle forecasting schemes \cite{Choudhurietal07,Jiangetal07,BhowmikNandy18}.
In such cases the details of the underlying flux transport dynamo model 
(Section \ref{sec:FTD})
are secondary, as long as shearing by differential rotation is linear,
i.e., there is no significant dynamical 
backreaction of the magnetic field on differential rotation,
and the associated inductive
shearing is not subjected to significant forcing by random fluctuations.

\begin{figure}[t]
\begin{center}
\includegraphics[width=0.8\linewidth]{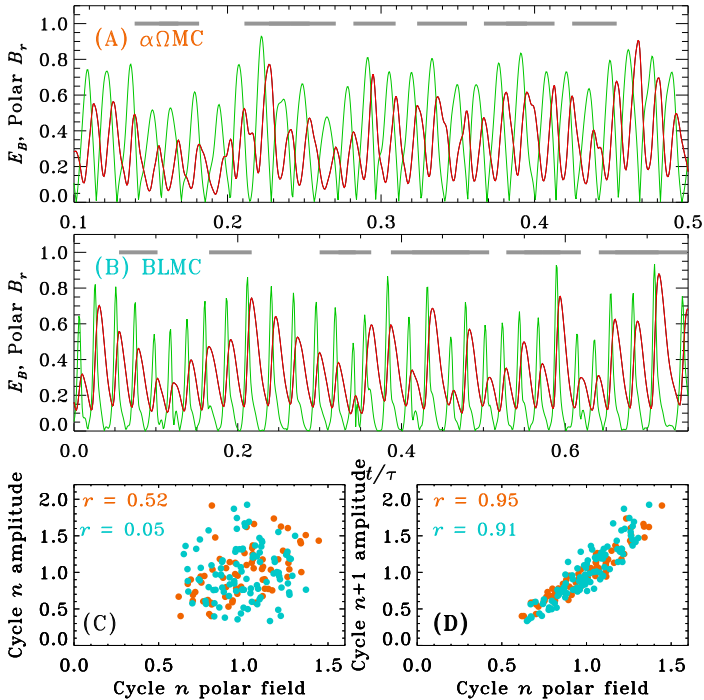}
\end{center}
\caption{
Two solar cycle-like solutions in flux transport
dynamo models differing only in their mechanism of poloidal field regeneration, and subjected to stochastic forcing of the latter (see text). Panel A and B
are respectively an
$\alpha\Omega$ and Babcock-Leighton solar cycle models, both including meridional circulation. 
Green lines are time series of the surface polar cap magnetic field, and red lines are time series of magnetic energy 
integrated over the solution domain, used here as a proxy of magnetic cycle amplitude. 
Neither model shows a significant
correlation between cycle amplitude and dipole strength at the subsequent minimum (panel C), but both show a strong correlation
between dipole strength at minimum and the amplitude of the subsequent cycle (panel D).
Figure 19 in \cite{Charbonneau20}, used with permission.}
\label{fig:stochanal}
\end{figure}

The good precursor value of the solar surface dipole is often matter-of-factly
invoked as empirical support for the Babcock-Leighton ```picture'' of the solar dynamo, i.e., the large-scale poloidal magnetic component being regenerated by the surface decay of bipolar
magnetic regions (viz.~Section \ref{sec:induction}). 
Figure \ref{fig:stochanal}, adapted from
\cite{CharbonneauBarlet11}, offers a specific counterexample to this claim. Panel (A) and (B)
show time series of volumetric magnetic energy (red) and surface dipole (green; more precisely: 
Northern hemisphere polar cap magnetic field) produced by two dynamo models differing only in their poloidal source; 
the solution of panel (A) is a conventional mean-field $\alpha\Omega$ dynamo model, with the $\alpha$-effect concentrated at the base of the convection zone, but includes a meridional flow and operates in the flux transport regime 
(Section \ref{sec:FTD}). 
The solution of panel B is a mean-field-like Babcock-Leighton dynamo model using a non-local surface poloidal source term, as described in \cite{DikpatiCharbonneau99}. 
Both models are axisymmetic, kinematic,
use the same solar-like parametrization of the solar internal differential rotation, 
and the quadrupolar meridional flow pattern of \cite{vanBallegooijenChoudhuri88}, 
characterized by a single flow cell per meridional quadrant spanning the full convection zone. 
In both cases zero-mean
stochastic fluctuations are imposed on the dynamo number multiplying the poloidal source term,
of amplitude corresponding to 50\% of the mean and with coherence time of one month.

As shown on panel C, in either model no correlation is observed between the dipole strength 
at minimum
and the amplitude of the cycle just ending, consistent with the fact that imposed random fluctuations affect the production of the poloidal component from the toroidal component. 
However, both models show a strong correlation between dipole strength and the amplitude of the subsequent cycle (panel D), as measured here via volumetric magnetic energy. 
In the case of the mean-field $\alpha\Omega$ model, this correlation vanishes altogether if the meridional flow is turned 
off and the model then operates as a classical $\alpha\Omega$ model, 
with equatorward propagation
of activity belts driven by a dynamo wave. 
The surface dipole then becomes a side-effect of a dynamo operating in the deep interior, and does not feed back into the dynamo loop. Nonetheless,
in the flux transport mean-field $\alpha\Omega$ model the surface dipole is as good a precursor of the next cycle as in the Babcock-Leighton model.

{\bf To sum up:} 
for the surface dipole moment to act as a good precursor of the upcoming cycle's amplitude, two conditions must be met: 
(1) the primary source of fluctuation must reside in the regeneration of the large-scale poloidal magnetic component, and 
(2) the surface dipole must feed back into the dynamo loop. Many dynamo scenarios
meet both constraints, 
and none can be favored over another only on the basis of the precursor value of the surface dipole.

\section{Tension: cycle fluctuations: stochastic or nonlinear?\label{sec:fluct}}

The sunspot numbers record reveals significant cycle-to-cycle variations in the amplitude, duration, shape, and hemispheric asymmetry of the solar cycle
(see \cite{Hathaway15}, and references therein). 
Reconstructions of solar activity based on cosmosgenic radioisotopes also reveals modulation patterns unfolding on centennial to millennial timescales \cite{Usoskin17}. What is the physical origin of these variability patterns ?

Solar and stellar dynamos draws their energy from the kinetic energy of the participating
inductive flows, through work done against Lorentz force associated with the dynamo-generated magnetic field. This is
a nonlinear backreaction of the magnetic field on the flows, 
which most certainly
is what prevents unbound growth of the cycle amplitude, and may also
drive amplitude modulation on timescale longer than the cycle period.

Solar and stellar dynamos
also operates in part or {\it in toto} in strongly turbulent convective envelopes. 
Turbulent convection acts as multiscale and spatiotemporally highly variable inductive component, 
which from the point of view of the large-scale magnetic field 
manifests itself as short coherence time stochastic ``noise'' superimposed
on the mean electromotive force and induction by large-scale flows.
Such stochastic noise can also cause significant cycle amplitude variability,
Figure \ref{fig:stochanal} herein being a case in point.

Is stochastic forcing or deterministic nonlinear backreaction driving observed cycle fluctuations ? 
This is a particularly complex question to answer because
in the solar/stellar context, many flow components contribute to induction 
(and/or flux transport), and
all are in principle impacted by the Lorentz force associated 
with the dynamo-generated large-scale magnetic field.
Moreover, the magnetic field can also impact large-scale flows indirectly,
via alterations of
of Reynolds stresses powering them (the so-called $\Lambda$-quenching; 
see \cite{Kitchatinovetal94,Kuekeretal96,Rempel06b}), 
or via global constraints such as magnetic helicity conservation, as discussed in Section \ref{sec:maghelicity}.
To further complicate matters, in general
the response of the dynamo to stochastic forcing depends on the nature 
of the nonlinearity regulating the average cycle amplitude (see, e.g.,
\cite{Biswasetal22,Talafhaetal22} and references therein).

Assorted dynamo modelling work 
(see, e.g., \cite{Mossetal92,Hoyngetal94,Ossendrijveretal96,Usoskinetal09a,
ChoudhuriKarak12,Pipinetal12,OlemskoyKitchatinov13,HazraNandy19} 
for a representative subset)
has amply demonstrated that in the presence of stochastic forcing,
many solar-like behaviors, such as marked amplitude fluctuations, sustained mixed-parity modes, cross-equatorial activity, and intermittency,
can materialize naturally in critical or very weakly supercritical dynamos.
R.~Cameron and M.~Sch\"ussler \cite{CameronSchuessler17} go further in arguing that a stochastically forced weakly supercritical dynamo is all that is needed to reproduce the observed spectral
properties of solar activity, in a manner that is generic with respect to the amplitude-limiting nonlinearity. 
The key of their proposal is that the linear dynamo
growth rate be much smaller than the cycle period ---as one would indeed expect for very weakly supercritical dynamos.
An ensemble average of their model runs
yields a flat spectrum at low frequencies, but any single instance 
is characterized by spectral structure, of purely random origin
(see their Fig.~3). 
Cosmogenic radioisotope reconstructions of solar activity on long timescale also show spectral structures at low frequencies, but also
represent a single realization of a specific dynamo ---the Sun! 
Can such reconstructions then actually prove or disprove the Cameron \&
Sch\"ussler conjecture ? 
Notwithstanding circumstantial evidence related to rotational
evolution \cite{KitchatinovNepomnyashchikh17,Metcalfeetal22}, 
there are no {\it a priori} reasons to believe that
the solar dynamo is only weakly supercritical.
Moreover, long timescale modulation of cyclic behavior, as well as
parity modulation, intermittency, etc, are also readily produced purely
deterministically through various nonlinearities, notably magnetic backreaction on differential rotation 
\cite{Tobiasetal95,Kuekeretal99,Pipin99,MossBrooke00,Bushby06,TobiasWeiss16,SimardCharbonneau20}
(see also Section 4 in \cite{Charbonneau20}).
In a picture purely based on stochastic forcing, one would not expect
any phase coherence in long term fluctuations. 
Cosmogenic reconstructions do suggest phase persistence and non-random clustering patterns for Grand Minima and Maxima (more on these further below), 
but even the most recent $9000\,$yr reconstructions \cite{Usoskinetal16,Wuetal18} 
remain too short to yield strong statistical significance to
confidently support or refute either class of explanation.

{\bf To sum up:} The exact nature of the magnetic nonlinear backreaction 
mechanism(s) stabilizing the amplitude of the solar dynamos is not
yet identified with confidence; nor are the mechanisms, whether of a stochastic
or purely deterministic nature, driving apparently random cycle-to-cycle
fluctuations in cycle characteristics such as amplitude and duration, as well
as variability on timescale longer than the cycle period.
These are absolutely fundamental gaps in our (lack of) understanding
of solar and stellar dynamos.

\section{Tension: Explaining Grand Minima\label{sec:Maunder}}
 
The most extreme pattern of solar fluctuation is arguably the Grand Minima
of solar activity, epochs during which activity falls to very low levels over
a time period much longer than the magnetic cycle period.
First noted in the sunspot record independently by G.~Sp\"orer and E.~W.~Maunder in the late nineteenth century, and much later rediscovered 
\cite{Eddy76}, the
1645--1715 Maunder Minimum has become the archetype of such events, 
which have been found to recur aperiodically in the cosmogenic radioisototope 
record spanning the Holocene
(see \cite{Usoskinetal15}, and references therein).
Relatively recent similar events include
the Sp\"orer Minimum (ca.~1416--1534) and
the Wolf Minimum (ca.~1282--1342). Other periods of sustained 
higher-than-average activity, ``Grand Maxima'', can also be identified
\cite{Usoskin17}. 

The Maunder Minimum remains unique in that it is the only Grand Minimum
for which direct observations of sunspots 
are available. Extensive historical
analyses have revealed that the Sun was not entirely devoid of sunspots
during the Maunder Minimum, but that the few sunspots observed were
almost all located in the Southern solar hemisphere 
\cite{RibesNesme-Ribes93,HoytSchatten96,Usoskinetal15}; 
for a comprehensive review
of historical sunspot observations, see \cite{ArltVaquero20}.
Another intriguing pattern relates to a possible change in the surface 
latitudinal rotation, as revealed from analyses of sunspot drawings made
before, during and after the Maunder Minimum 
\cite{Eddyetal77,Casasetal06}.

Dearth in production of sunspot does not necessarily mean a halt in
cyclic regeneration of the solar large-scale magnetic field. The buoyant
destabilization
of magnetic flux rope formed in the deep solar interior (presumably), 
with subsequent rise through the convective envelope and emergence as
bipolar magnetic regions, almost certainly involves a threshold in
magnetic intensity \cite{Schuessleretal94}. Consequently, 
the magnetic cycle may well
continue unabated through Grand Minima, without reaching a magnetic amplitude
sufficiently high to produce sunspots. Under this view, Grand Minima
simply  represent
the low point of a large amplitude modulation. 
It is noteworthy that determinisitc, 
nonlinearly-driven backreaction on large-scale inductive flows 
(Section \ref{sec:fluct}) can be accompanied
by modulation of both flow and magnetic equatorial
parity unfolding on long timescales, thus offering an attractive explanation
for the 
strong hemispheric asymmetry of sunspot locations during the Maunder Minimum
\cite{SokoloffNesme-Ribes94,Beeretal98}, as well as any
variation in differential rotation. For a sample of dynamo models
exhibiting this type of nonlinear modulation, 
see \cite{Kuekeretal99,Bushby06,WeissTobias16}

A distinct class of explanations for Grand Minima invokes \emph{intermittency}
\cite{Plattetal93},
namely a transition between two distinct dynamo modes, the weaker one non-cyclic
and/or of very low magnetic amplitude. 
The switch between modes can be driven either stochastically or deterministically. For a sample of dynamo models generating Grand Minima-like episodes in this
manner, see
\cite{Ossendrijver00a,Charbonneauetal04,Mossetal08,Usoskinetal09a,ChoudhuriKarak12,OlemskoyKitchatinov13,Inceogluetal17,Albertetal21}.
Explaining Grand Minima via intermittency does pose a problem for dynamo models 
which are not self-excited, in the sense of being subjected to a lower operating threshold, 
e.g., dynamos relying on the Babcock-Leighton mechanism 
(Section \ref{sec:induction}).
Jump-starting
the primary dynamo out of a Grand Minimum
then requires either a secondary self-excited dynamo
\cite{Passosetal14,Hazraetal14b,Olceketal19,Sahaetal22}, 
or some other source of magnetic
fields acting as ``magnetic noise'' (see, e.g.,
\cite{Schmittetal96,Ossendrijver00a,Charbonneauetal04,Tripathietal21}).

The observations of low amplitude cyclic activity in the ${}^{10}$Be isotope
record has been presented as evidence that the solar cycle was still running
throughout the Maunder Minimum \cite{Beeretal98},
the interplanetary magnetic field being a complex sampling of both active region magnetic fields as well as the global dipole. A turbulent $\alpha\Omega$
dynamo may well continue to reverse polarity, while failing to reach a
magnetic amplitude sufficient to generate large
emerging BMRs and associated sunspots. This type of behavior does materialize
more naturally in dynamos undergoing large amplitude modulation through 
nonlinear backreaction by the Lorentz force.
However,
in some flux transport models undergoing intermittency, 
cyclic variations of the 
surface field can also take place as the meridional flow entrains the residual
magnetic field (see \cite{Charbonneauetal04,Sahaetal22} for specific examples).
The persistence of cycle \emph{phase} through Grand Minima is a potentially
powerful discriminant, but quite challenging to harness in practice.

{\bf To sum up:} 
A wide variety of potentially viable dynamo-based scenarios for Grand Minima have been proposed, but which (if any) actually applies to the Sun remains an open
question.

\section{Tension: Sensitive cycles in MHD simulations\label{sec:MHDsims}}

For now more than a decade, many global magnetohydrodynamical simulations
of solar convection have achieved the production of large-scale magnetic fields undergoing more or less regular polarity reversals, analogous to some extent
to the solar magnetic cycle. For a representative sample of such simulations, see, e.g.,
\cite{Ghizaruetal10,Masadaetal13,Nelsonetal13,FanFang14,Mabuchietal15,Simitevetal15,Hottaetal16,Kapylaetal17,Strugareketal18,Guerreroetal19,Hottaetal22}.
These simulations rely on markedly distinct computational
approaches to the numerical solution of the MHD equations,
in particular in the treatment of unresolved scales. 
While they often generate similar convective and large-scale flow patterns, 
the large-scale magnetic cycle they produce vary widely in their spatiotemporal evolution (see Section 3.2 in \cite{Charbonneau14} for a specific comparison).
The origin of this ``structural fragility'' is
multi-faceted and remains ill-understood.

\begin{figure}[ht]
\begin{center}
\includegraphics[width=0.6\linewidth]{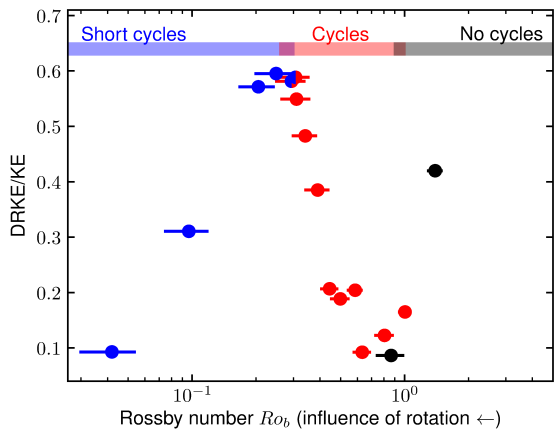}
\end{center}
\caption{
A synthetic summary of a set of simulations from \cite{Strugareketal18}, showing
the ratio of kinetic energy contained in differential rotation to total kinetic energy, versus Rossby number. Deep seated, solar-like decadal magnetic cycles 
are plotted in red. The blue points refer to short (period of a few years) magnetic cycles unfolding in the upper third of the convecting fluid layers, while black point indicate simulations generating steady large-scale magnetic fields.
Reproduced with permission from \cite{Strugareketal18}, copyright by AAS.}
\label{fig:summary}
\end{figure}

Figure \ref{fig:summary} summarizes cycle properties
in a series of global MHD simulations from \cite{Strugareketal18}, collectively
spanning a factor of 10 in rotation rate and 3 in convective luminosity. 
The ratio of kinetic energy contained in differential rotation (DRKE) 
to total kinetic energy (KE) is plotted 
against Rossby number, with symbols colored according to the character
of the large-scale magnetic cycle materializing in each simulation.

The sensitivity on the Rossby number is extreme; starting at $\Ro\simeq 0.1$,
increasing $\Ro$ by a mere a factor 
of 10 takes one from rapid subsurface cycles, through deep-seated decadal 
cycles in the range $0.3\lta\Ro\lta 1$, 
and on to steady large-scale magnetic field at $\Ro\gta 1$.
This extreme sensitivity turns out to be robust, in the sense that it materializes in MHD simulations using entirely distinct numerical implementations
and overall modelling frameworks; consider the striking resemblance between
Figure 13 in \cite{Brunetal22}, working with the ASH LES code 
\cite{Brunetal04}, with
Figure \ref{fig:summary} herein, from \cite{Strugareketal18} using
the EULAG-MHD ILES code \cite{SmolarkiewiczCharbonneau13}.

There is
much more to the sensitivity issue than the Rossby number, however.
The large-scale flow and magnetic field emerging in MHD numerical simulations of solar convection are strongly influenced by the
turbulent regime attained. At high Reynolds number, stresses associated with the turbulent magnetic field can have a strong impact on large-scale flows
\cite{FanFang14,HottaKusano21,Brunetal22,Hottaetal22}, thus indirectly affecting global dynamo action. 
Likewise, the relative importance of the many dissipation channels available 
to the system, as measured by the viscous and magnetic Prandtl numbers, 
also influences significantly the large-scale flows and dynamo action \cite{Kapylaetal17,Tobias21}.

Not surprisingly,
the numerical treatment of small scales also impacts turbulent induction.
Mean-field analyses of large-scale magnetic cycles in MHD simulations
typically yield $\alpha$-tensors that are full, 
with turbulent induction often contributing on par with large-scale shearing
\cite{Augustsonetal15,Simardetal16,Warneckeetal18,Vivianietal19}.
In mean-field parlance, these simulations operate as $\alpha^2\Omega$ dynamos,
or even $\alpha^2$ dynamos, if turbulent induction dominates over shearing
by differential rotation. Detailed analyses of various simulations also reveal that small-scale and large-scale inductive contribution sometimes counteract each other \cite{Racineetal11,Nelsonetal13,Brunetal22,Shimadaetal22}, which results
in the total induction having a magnitude significantly smaller 
than its individual contributions. Relatively small changes in one contribution, for example in small-scale induction versus dissipation
when distinct subgrid models are used or when the magnetic Prandtl number is varied,
can have a much larger impact on total induction, and thus on the unfolding of
the large-scale magnetic cycle. 

Another complication arises in MHD simulations reaching very high Reynolds numbers \cite{Hottaetal16}, namely small-scale dynamo action. 
Producing a strong small-scale magnetic field in this manner turns out
to suppress the small-scale turbulent flow 
otherwise responsible for the turbulent diffusion
of the large-scale magnetic component \cite{Shimadaetal22}. 
Somewhat counterintuitively, more strongly turbulent simulations end up
sustaining large-scale magnetic fields better than less turbulent ---and presumably less diffusive--- simulations. 
Here again the computational treatment of subgrid effects can have a large impact, and the key to stable global magnetic cycles appears to be the minimization
of dissipation at the larger scales \cite{Strugareketal16,Strugareketal18}. 

Finally, the combination of multiple turbulent inductive contributions and relatively complex internal differential rotation profiles with distinct shearing regions can lead to the co-existence of multiple dynamo modes, spatially segregated but nonetheless interfering with one another, generating variability and
modulation of magnetic cycle unfolding in MHD simulations \cite{Beaudoinetal16,Kapylaetal16,Strugareketal18,Vivianietal19}. Interference between distinct dynamo modes can also
yield occasional episodes of strongly reduced cyclic activity
\cite{Augustsonetal15,Kapylaetal16},
somewhat reminiscent of Grand Minima (viz.~\S 9).

{\bf To sum up:} Global MHD numerical simulations of solar convection can generate large-scale magnetic fields undergoing more or less regular polarity reversals. The unfolding of these magnetic cycles seems however far more sensitive to modelling details and physical parameter regimes that suggested by observations
of magnetism and cycles in solar-type stars. The physical
origin of this sensitivity remains an open question.

\section{Tension(s): From solar to stellar dynamos\label{sec:soltostar}}


The sun is but a star, yet its ease of observation makes it an essential springboard towards interpreting magnetic activity and cycles observed in other stars
as resulting from the operation of dynamos. 
This is an immense topic, and we will close this review by only highlighting a few key points.

As revealed by the epoch-making Mt Wilson Ca H$+$K survey 
\cite{Baliunasetal95}, all solar-type stars show evidence of magnetic activity,
and a subset shows fairly well-defined cyclic variability on decadal timescales,
presumably reflecting the presence of a dynamo-powered 
large-scale magnetic cycle analogous to the solar cycle.
Observation of X-Ray emission, a tracer of coronal activity powered by magnetism, show a well-defined variation with the Rossby number inferred 
from the observed luminosity via mixing length theory. Particularly noteworthy
is the fact that this trend is the same for solar-type stars (meaning,
G and K dwarfs having a radiative core and overlying 
convective envelope), fully convective M dwarfs \cite{Wrightetal18}, 
and even
evolved subgiants and giants \cite{Lehtinenetal20}. This 
suggests ---without strictly proving---
convective turbulence-related universality in
the dynamo mechanism underlying stellar magnetic activity at large.

Again in late-type main-sequence stars,
measurements of surface magnetism also show a fairly well-defined trend with
Rossby number, also largely independent of spectral type
\cite{Reinersetal22}. This points again to a certain level of universality
in stellar dynamo, which is not obvious to reconcile with the
many detailed dynamo scenarios designed specifically
to match solar observations (but do see \cite{Shulyaketal15}).

Even if complete knowledge of the solar dynamo were at hand ---and it is not, as pointed out throughout this review,---
going to the stars (even ``only'' to late-type main-sequence stars)
demands answers to a number of crucial questions, 
including minimally:

\begin{enumerate}
\item Which is the primary polarity reversal mechanism: $\alpha$-effect, or Babcock-Leighton,... or something else ?
\item How do differential rotation and meridional circulation
vary with rotation rate, luminosity, and internal structure ?
\item How do turbulent coefficients ($\alpha$-effect, turbulent pumping,
turbulent diffusion) vary with rotation rate, luminosity, 
and internal structure ?
\item How do sunspots and BMRs form and decay in stars of varying structure
(in particularly, depth of convective envelope), 
rotation rate and luminosity ?
\end{enumerate}

Harnessing knowledge acquired via solar dynamo modelling
and MHD numerical simulations can in principle adress many of these questions;
for a thorough review see \cite{BrunBrowning17}. The variation of differential
rotation and meridional circulation is accessible via both semi-analytical turbulence models (e.g., \cite{KitchatinovRuediger93}) and numerical simulations
(\cite{Brunetal22}, and references therein), and there is now general agreement
that (latitudinal) differential rotation does not vary much with rotation
rate once in the solar-like (rapidly rotating equator) regime. 

The behavior of large-scale
magnetic cycles, on the other hand, shows greater disagreement between
observations and theory; as a single example, consider the variation of the cycle period with Rossby number: the original dynamo analysis of Mt Wilson data by
\cite{Noyesetal84} suggested
$P_{\rm cyc}/P_{\rm rot}\propto \Ro^{0.25}$, while 
the distinct numerical simulations of \cite{Strugareketal18} and \cite{Warnecke18}
indicate 
$P_{\rm cyc}/P_{\rm rot}\propto \Ro^{-1.6}$ and $\Ro^{-1.8}$, respectively. 
Reliably estimating
the Rossby number is far from trivial, either from stellar data or from
numerical simulations
(see, e.g., \cite{Brunetal22}), but the fact that these 
two trends run in opposite direction is worth reflecting upon, to say the least.

{\bf To sum up:} 
The rapidly growing body of high-quality observations of stellar activity
and magnetism begs for the design of a unifying dynamo framework applicable
to both the sun and solar type stars of varying spectral type, luminosity, 
and rotation rate. Which are the key elements on which to build such a framework remains an open question.

\bmhead{Acknowledgments}

This review was written following the workshop
``Solar and Stellar Dynamos: A New Era'', hosted and supported by the International Space Science Institute (ISSI) in Bern, Switzerland. The author wish to express their thanks to
ISSI for their financial and logistical support.

\bmhead{Declarations}

The authors have no relevant financial or non-financial interests to disclose.

\def\araa{{\it Ann.~Rev.~Astron.~Ap.}}
\def\aap{{\it Astron.~Ap.}}
\def\apj{{\it Astrophys.~J.}}
\def\apjl{{\it Astrophys.~J.~Lett.}}
\def\mnras{{\it Month.~Not.~Astron.~Soc.}}
\def\solphys{{\it Solar Phys.}}
\def\ssr{{\it Sp.~Sci.~Rev.}}
\def\prl{{\it Phys..~Rev.~Lett.}}
\def\grl{{\it Geophys.~Res.~Lett.}}
\def\jgr{{\it J.~Geophys.~Res.}}

\bibliographystyle{plain}
\bibliography{refs-issi2022}

\end{document}